\documentclass[12pt]{iopart}
\usepackage{graphicx}
\usepackage{amssymb}

\begin{document}


\title{Lorentz-violation
and cosmological perturbations: a toy brane-world model}

\author{M.V.~Libanov\dag\ and V.A.~Rubakov\dag}
\address{\dag\
Institute for Nuclear Research
of the Russian Academy of Sciences,\\
  60th October Anniversary prospect 7a, Moscow 117312, Russia}
\eads{\mailto{rubakov@ms2.inr.ac.ru}, \mailto{ml@ms2.inr.ac.ru}}

\begin{abstract}
We study possible effects of Lorentz-violation on the generation of
cosmological perturbations at inflation by introducing a simple inflating
five-dimensional brane-world setup with violation of four-dimensional
Lorentz-invariance at an energy scale $k$. We consider massless scalar
field, meant to mimic perturbations of inflaton and/or gravitational
field, in this background. At three-momenta below $k$, there exists a zero
mode localized on the brane, whose behaviour coincides with that in
four-dimensional theory. On the contrary, at three-momenta above $k$, the
localized mode is absent and physics is entirely five-dimensional. As
three-momenta get redshifted, more modes get localized on the brane, the
phenomenon analogous to ``mode generation''. We find that for $k\gg H$,
where $H$ is the inflationary Hubble scale, the spectrum of perturbations
coincides with that in four-dimensional theory. For $k < H$ and
time-dependent bulk parameters, the spectrum deviates, possibly strongly,
from the flat spectrum even for pure de Sitter inflation.
\end{abstract}

\pacs{
98.80.-k,
04.50.+h,
}
\maketitle
\section{Introduction and summary}
\label{sec:intro}

The standard inflationary mechanism of the generation of cosmological
perturbations  assumes Lorentz-invariance. This assumption is worth
questioning~\cite{d1,d2,d3,d4,d5,d6,d7,v1,v2,v3,v4,m1,m2,
m3,m4,m5,g1,g2,nc2,br1,br2,br3} especially in view of the fact that
relevant perturbations initially have wavelengths much shorter than the
Planck length (this observation is often referred to as ``trans-Planckian
problem''). Furthermore, there is no guarantee that Lorentz-invariance
holds even towards the end of inflation. Thus, it is certainly of interest
to understand the effects that Lorentz-violation may have on the
cosmological perturbations.

Several ways to address this issue have been proposed.  One is to assume
Lorentz-violating dispersion relation at short enough
wavelengths~\cite{d1,d2,d3,d4,d5,d6,d7}. If at asymptotically high momenta
the frequencies exceed the inflationary Hubble parameter, the adiabatic
vacuum is a natural choice of the initial state, and then the primordial
spectrum is still flat for pure de~Sitter inflation~\cite{d1,d2,d3,d4,d5},
while the predictions for the tilt may be different, as compared to the
standard mechanism, if the expansion is not exactly de~Sitter. The
opposite case of low frequencies at high momenta~\cite{d1,d2} is hard to
treat: perturbations are not initially in the adiabatic regime, so there
is no preferred initial state (an exception here is the bouncing Universe
model of reference~\cite{d6}).

Another approach is to make ad hoc assumptions on the initial state of
perturbations~\cite{v1,v2,v3,v4}. The problem with this approach is that
any choice other than the adiabatic vacuum is hard to justify, whereas
different choices lead to quite different spectra of
perturbations~\cite{v3,v4}.

One more possibility has to do with ``mode generation'': there are models
where the number of modes increases in time, as the Universe
expands~\cite{m1,m2,m3,m4,m5}; this happens, e.g., if the physical
momentum is effectively bounded from above. This option is in a sense
similar to the previous one: the result for the spectrum of perturbations
strongly depends on the choice of a state in which a newborn mode emerges.
If this is the adiabatic vacuum, the outcome is similar to the standard
scenario, at least for pure de~Sitter inflation~\cite{m2,m3,m5}.

The emerging picture is that in the least exotic scenarios, the spectrum
of perturbations is flat for pure de~Sitter inflation, while predictions
for the tilt in non-de~Sitter cases are generally different from those of
the standard mechanism. If the energy scale $k$ associated with
Lorentz-violation is much higher than the Hubble parameter $H$ at
inflation, the deviations of the spectrum from the predictions of the
standard mechanism are suppressed by powers of $H/k$, as argued on general
grounds, e.g., in references~\cite{d4,g1,g2}. On the other hand, dramatic
assumptions, like, e.g., the Corley--Jacobson dispersion
relation~\cite{cj} with frequencies tending to zero at high momenta, lead
to dramatic consequences: the spectrum may be quite different from the
flat spectrum even for pure de~Sitter inflation~\cite{d1,d2,d7}. Also, for
$k\lesssim H $ the properties of the primordial perturbations may deviate
substantially from the standard predictions even in some less exotic
models, like, e.g., the model of reference~\cite{nc2} based on
noncommutative field theory.

Violation of four-dimensional Lorentz-invariance is fairly natural in
brane-world models~\cite{visser,freese,kolb,csaki,sergd}. It may occur in
the bulk, especially if there is bulk matter (a particularly
well-understood solution of this sort is a  brane moving in
Schwarzschild--anti-de~Sitter bulk~\cite{krauss,gregory}) or
Lorentz-violating bulk condensate~\cite{sibir}. Since the wave functions
of brane matter may extend to the bulk, the dispersion relations on the
brane are then naturally Lorentz-violating~\cite{csaki}. Furthermore, some
brane-world geometries have the property of ``mode generation'' in the
following sense: for low three-momenta, there exist modes localized (or
quasi-localized) on the brane, while for high three-momenta, localized and
even quasi-localized modes are absent~\cite{sergd}. Thus, as the Universe
expands and three-momenta get redshifted, more and more brane modes
emerge; this may be viewed as the brane-world version of mode generation.
Yet the initial state is well defined, provided that modes with high
three-momenta are adiabatic in the bulk.

In this paper we consider an inflating version of the setup of
reference~\cite{sergd}, and study massless scalar field, meant to mimic
perturbations of the inflaton and/or gravitational field, in this
background. This is nothing more than a toy model, as we specify neither
the mechanism giving rise to Lorentz-violation, nor the mechanism driving
inflation. In fact, our background may be totally unrealistic, as it may
require matter with unphysical properties either in the bulk or on the
brane, or both. Also, since we do not have a mechanism generating our
background, we cannot quantitatively address the issue of back
reaction~\cite{br1,d4,br2,br3} of the perturbations on background geometry
(however, we will argue on qualitative grounds that this back reaction is
small). On the other hand, the advantage of our approach is that the
properties of the massless scalar field are well understood in spite of
mode generation on the brane and other unusual features.

To a certain extent we confirm the picture emerging from the previous
studies. In the simplest version of our setup, with time-independent bulk
parameters, the spectrum of perturbations generated due to inflation is
flat for pure de~Sitter expansion, while its amplitude and, in the case of
non-de~Sitter expansion, the tilt are generally different from those
predicted by the standard mechanism. Furthermore, for $k \gg H$, where $k$
and $H$ are still the scales associated with Lorentz-violation and
expansion rate at inflation, respectively, even the latter deviations are
small.

\begin{figure}[b]
\unitlength=1mm
\begin{center}
\begin{picture}(155.00,70.00)(-30,5)
\put(-8,60){\mbox{$\mathcal{P}(p)$}}
\put(-8,40){\mbox{$\epsilon ^{-2\alpha }H^2$}}
\put(-1,13){\mbox{$ H^2$}}
\put(93,2){\mbox{$p$}}
\put(58,2){\mbox{$\epsilon ^{-1}p_\times $}}
\put(31,2){\mbox{$p_\times $}}
\includegraphics[height=70mm]{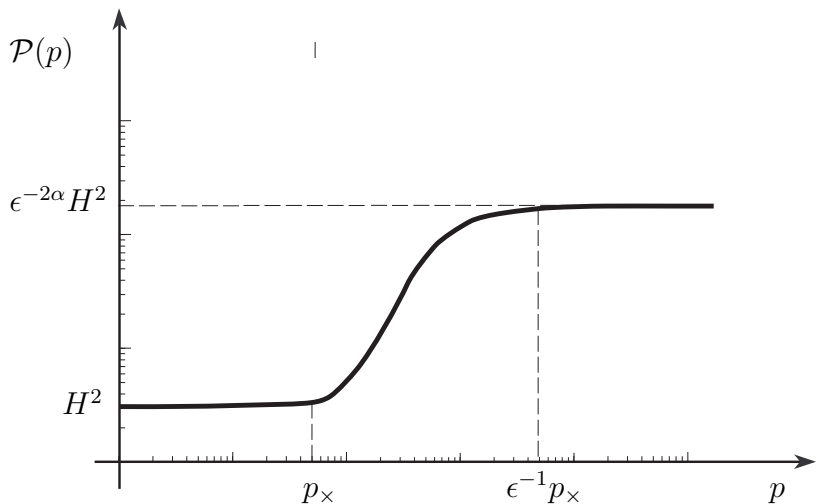}
\end{picture}
\end{center}
\caption{\label{figure1} Power of perturbations, as function of
momentum $p$, generated in a model with $k$ decreasing in time.
$p_\times$ is a free parameter and $\epsilon \ll 1$ is another free
parameter of the model. The exponent $\alpha $ is model dependent and
varies in the range $0<\alpha <3/2$. See section~\ref{decaying k} for
explanation and qualifications.}
\end{figure}

Things are more interesting if one makes a relatively mild assumption that
the bulk parameters depend on time, so that $k$ is smaller than $H$ towards
the end of inflation. As an example, we consider the case in which $k$
slowly decreases in time, from $k\gg H$ at the early stage of inflation to
$k<H$ towards its end. Then the primordial spectrum is flat at low momenta,
gets strongly tilted at intermediate momenta, and flattens out again at
high momenta, as sketched in figure~\ref{figure1}. The position of the
tilted part, as well as its width, depend on the parameters of the model;
this part may well extend over several (but not necessarily many)
decimal places, and the amplitudes at low and high momenta may differ also
by several orders of magnitude. To avoid confusion, the primordial
spectrum is cut off in the ultraviolet in the usual way, the cutoff being
at physical momenta of order $H$ at the end of inflation.

\begin{figure}[b]
\unitlength=1mm
\begin{center}
\begin{picture}(155.00,70.00)
(-30,5)
\put(-8,60){\mbox{$\mathcal{P}(p)$}}
\put(-8,40){\mbox{$\epsilon ^{-2\alpha }H^2$}}
\put(-1,13){\mbox{$ H^2$}}
\put(93,2){\mbox{$p$}}
\put(59,2){\mbox{$p_\times $}}
\put(29,2){\mbox{$\epsilon p_\times $}}
\includegraphics[height=70mm]{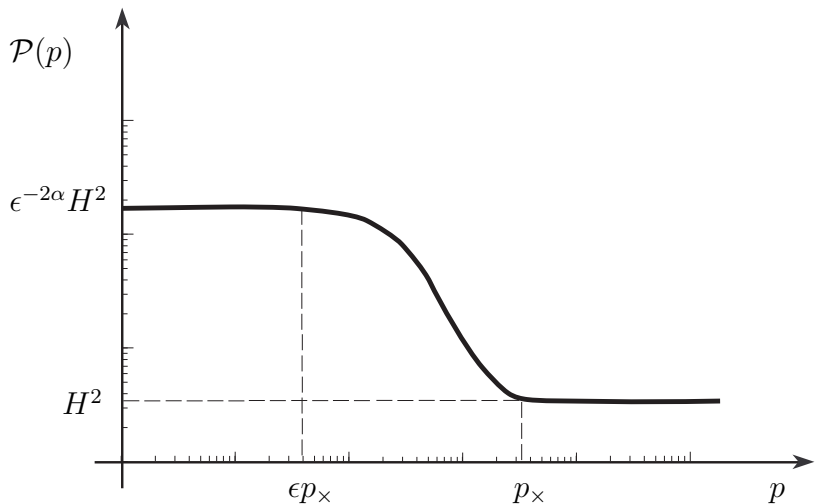}
\end{picture}
\end{center}
\caption{\label{figure2}
Same as in figure~1 but with $k$ increasing in time.}
\end{figure}

The opposite situation, shown in figure~\ref{figure2}, occurs if $k$
increases in time from $k < H$ to $k\gg H$.

A lesson we learn from our toy model is twofold. On the one hand,
we confirm that
it is quite unlikely that in a well defined setup with
Lorentz-violation well above the inflationary Hubble scale ($k \gg
H$ in our notations), the standard predictions may be strongly
modified. In our model, physics at three-momenta above $k$ is
entirely multi-dimensional, and yet for $k \gg H$ this has
basically no effect on primordial perturbations. On the other
hand, if Lorentz-violation is characterized by low energy scale $k
\lesssim H$, the spectrum of perturbations strongly depends on details
of the cosmological evolution at inflation, and may well be
grossly different from the standard approximately flat spectrum.

The spectrum sketched in figure~\ref{figure1} is interesting. If our
scalar field is interpreted as mimicking tensor perturbations,
this spectrum corresponds to greater (maybe much greater)
primordial amplitudes of gravity waves at shorter
wavelengths\footnote{As usual, the gravity waves decay as $a^{-1}$
after re-entering the horizon.}. If so, a possible non-observation
of the effects of tensor perturbations on CMB would not
preclude the detection of primordial gravity waves at shorter
wavelengths by techniques ranging from timing of pulsars to
ground-based interferometers. Moreover, since the latter
techniques cover a wide range of wavelengths, one might be able,
with luck, to uncover an interesting pattern of relic
stochastic gravity
waves in our Universe.

Likewise, the spectrum shown in figure~\ref{figure2} would correspond
to relatively high power at long wavelengths, possibly measurable
through CMB, with deviations from flatness potentially detectable
by experiments searching for gravity waves at shorter wavelengths.

We conclude this section by mentioning that even more complicated
spectra are obtained if one allows for time-dependence not only of
$k$, but also of other bulk parameters. We will comment on this
point in section~\ref{sec:timedependent}.

\section{Setup}
\label{sec:setup}

We consider $(4+1)$-dimensional model and denote the coordinates
as $(\eta, x^i,y)$, $i=1,2,3$. We take the five-dimensional metric
in the following form (signature $+,-,-,-,-$),
\begin{equation}
ds^2 =a^2(\eta)[\alpha^2(y) d\eta^2 - \beta^2(y) d{\bf x}^2] -
\alpha^2 (y) dy^2
\label{1.1*}
\end{equation}
where we made a convenient choice of the coordinate $y$. There is
a single brane at
\[
y=y_b =0
\]
As usual, the warp factors $\alpha (y)$ and $\beta (y)$ are continuous
across the brane, while their derivatives $\partial _y\alpha $ and
$\partial _y\beta $ are not. For equal warp factors, $\alpha (y) = \beta
(y)$, and static geometry $a(\eta) = \mbox{const}$, equation~(\ref{1.1*})
has a general form of warped metric invariant under four-dimensional
Lorentz group. For $\alpha (y) \neq \beta(y)$, and still $a(\eta
)=\mbox{const}$, the coordinates can be chosen in such a way
that
the induced metric on the brane is still
four-dimensionally Lorentz-invariant, while the full five-dimensional
metric is not. This is the case studied, e.g., in references~\cite{csaki,
sergd}. In what follows we choose coordinates so that the brane is
Lorentz-invariant for $a=\mbox{const}$. To specify our setup further, we
assume that both $\alpha(y)$ and $\beta (y)$ are $Z_2$-symmetric across
the brane and decay away from the brane,
\[
\alpha(y), \beta(y) \to 0 \; , \;\;\;\; \mbox{as} \; y\to \infty
\]
and that their ratio tends to a small constant,
\begin{equation}
\frac{\alpha(y)}{\beta(y)} \to \epsilon \; , \;\;\;\; \mbox{as} \; y\to
\infty
\label{1.1++}
\end{equation}
\[
\epsilon \ll 1
\]
The case $\epsilon=0$ and $a(\eta)=\mbox{const}$ was discussed in
reference~\cite{sergd}.

We are interested in inflating setup. For de~Sitter inflation, which is
the main focus in this paper, one has
\begin{equation}
a(\eta)= -\frac{1}{H\eta}
\label{1.1+}
\end{equation}
If inflation is not exactly de~Sitter, the Hubble parameter $H$ slowly
varies in time. Later on we will also consider generalizations of the
setup (\ref{1.1*}) in which $\alpha$ and/or $\beta$ slowly vary in time as
well.

Throughout this paper we concentrate on the properties of
five-dimensional massless minimal scalar field $\phi$ in the
background (\ref{1.1*}). Its action is
\[
S_\phi = \frac{1}{2} \int~d^5X~ \sqrt{g} g^{AB} \partial_A \phi
\partial_B \phi
\]
Upon introducing another field $\chi $ via
\begin{equation}
\phi = \frac{1}{a \beta^{3/2}} \chi
\label{1.2*}
\end{equation}
the action becomes
\[
\fl S_\chi = \frac{1}{2} \int~d\eta d^3x dy \left \{ \dot{\chi}^2
+\frac{\ddot{a}}{a} \chi^2 - \frac{\alpha^2(y)}{\beta^2
(y)}(\partial_{\bf x} \chi)^2 -
a^2 (\eta) \left[ (\chi^\prime)^2
+ \left(\frac{3}{2}\frac{\beta^{\prime\prime}}{\beta} +\frac{3}{4}
\frac{\beta^{\prime 2}}{\beta^2}\right) \chi^2 \right] \right \}
\]
where dot and prime denote $\partial_\eta$ and $\partial_y$,
respectively. The field equation in terms of three-dimensional
Fourier harmonics is
\begin{equation}
- \ddot{\chi} + \frac{\ddot{a}}{a} \chi - U(y) p^2 \chi -
a^2(\eta)\left[ -\chi^{\prime \prime} + V(y) \chi \right] = 0
\label{7}
\end{equation}
where $p$ is conformal three-momentum, and the ``potentials'' are
\[
   U(y) = \frac{\alpha^2(y)}{\beta^2(y)}
\]
and
\[
V(y) = \frac{3}{2}\frac{\beta^{\prime \prime}}{\beta} +
\frac{3}{4}\frac{\beta^{\prime 2}}{\beta^2}
\]
In what follows we assume that $V(y)$ tends to a constant as $|y|
\to \infty$, which, for future convenience, we denote as
\[
 V(|y| \to \infty) = \frac{9}{4} k^2
\]
$U(y)$ is a monotonically decreasing function whose
asymptotics, according to equation~(\ref{1.1++}), is
\[
U(|y| \to \infty) = \epsilon^2
\]
One more assumption is that the width $y_0$
of the ``potential'' $U(y)$,
i.e., the size of the region where $U(y)$ substantially exceeds
$\epsilon^2$, is sufficiently small; roughly speaking,
\begin{equation}
   y_0 < k^{-1}
\label{m6*}
\end{equation}
The reason for this assumption will become clear later\footnote{We
are grateful to W.~Unruh who pointed out the relevance of this
assumption, which simplifies the analysis considerably. We expect,
however, that our main results remain valid at $y_0 \gtrsim k^{-1}$
and even
$y_0 \gg k^{-1}$. We do not attempt to perform the anaysis in the
latter cases, as our attitude in this paper is ``proof by example''.}.
Finally, we will assume that $k$ and $\epsilon$ are the only relevant
parameters entering $U(y)$ and $V(y)$, the width $y_0$ being roughly of
order $k^{-1}$ or maybe substantially smaller.

We have to specify the boundary condition on the brane. We impose
$Z_2$ symmetry across the brane, consider the theory on half-space
$y \geq 0$, and, guided by the properties of
gravitational perturbations in the Randall--Sundrum model (see,
e.g.,~\cite{BC} and references therein), impose the condition
\[
 \phi^\prime (y=0) = 0
 \]
 i.e.,
\begin{equation}
 \left( \chi^\prime - \frac{3}{2} \frac{\beta^\prime}{\beta}
 \chi\right)_{y=0} = 0
 \label{print7*}
\end{equation}
This is in accord with the discontinuity of $\beta '(y)$ at $y=0$, which
results in the $\delta $-function behaviour of $V(y)$,
\begin{equation}
V(y)\approx 3\frac{\beta '}{\beta }\delta (y)\ \ \ \mbox{at}\  \ y\approx 0
\label{Eq/Pg8/1:paper}
\end{equation}
Hereafter we assume that $\beta^\prime <0$ for all $y$, so that this
contribution to the potential is attractive.

Although our analysis will not depend on the precise form of
the warp factors $\alpha (y)$ and $\beta (y)$, in
section~\ref{sec:example}
we will give an explicit
example by choosing
\begin{equation}
   \beta (y) = \mbox{e}^{-ky}
\label{m7+}
\end{equation}
Then
\[
  V(y) = \frac{9}{4} k^2 = \mbox{const}
\]
Equation~(\ref{7}) is simplified further for a particular choice
\begin{equation}
   U(y) = \frac{2}{3k} \delta (y) + \epsilon^2
\label{8**}
\end{equation}
(the reason for the coefficient $2/(3k)$ here will become clear later).
With this choice, the field equation and boundary condition read
\begin{equation}
- \ddot{\chi} + \frac{\ddot{a}}{a}\chi  - \epsilon^2 p^2\chi  -
a^2 (\eta) \left[ -\chi^{\prime \prime} + \frac{9}{4} k^2 \chi
\right] = 0
\label{8*}
\end{equation}
\begin{equation}
\left( \chi^\prime + \frac{3}{2} k \chi - \frac{1}{3k}
 \frac{p^2}{a^2 (\eta)} \chi \right)_{y=0} = 0
\label{8+}
\end{equation}
Note that even though variables separate in the field equation
(\ref{8*}), the dynamics is non-trivial because of the
time-dependence of the boundary condition (\ref{8+}).

\section{Static background}

\begin{figure}[htb]
\unitlength=1mm
\begin{center}
\begin{picture}(155.00,70.00)(-30,5)
\put(28,65){\mbox{$W_p(y)$}}
\put(73,25){\mbox{$y$}}
\put(53,57){\mbox{$p^2+O(k^2)$}}
\put(53,32){\mbox{$\epsilon ^2p^2+\frac{9}{4}k^2$}}
\includegraphics[height=70mm]{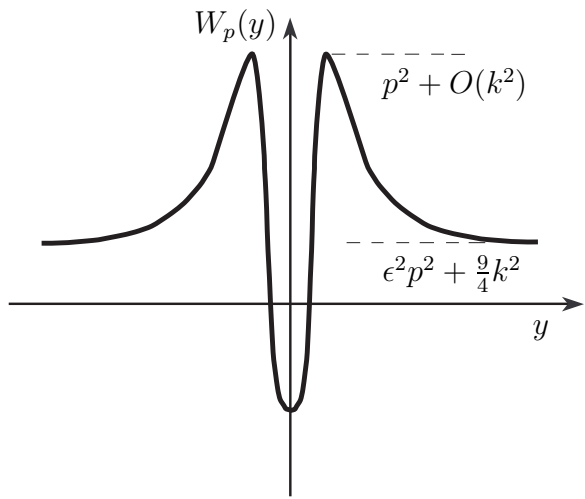}
\end{picture}
\end{center}
\caption{\label{figure3} Volcano-shaped effective potential
$W_p$. Its height and large-$y$ asymptotics increase with $p$,
while the dip at $y=0$ is independent of $p$.}
\end{figure}

In static background, $a(\eta) = 1$, and with
$\chi = \mbox{e}^{-i\omega \eta} \chi_\omega$, equation (\ref{7})
and boundary condition (\ref{print7*}) take the form of the static
Schr\"odinger equation
\begin{equation}
- \chi_\omega^{\prime \prime} + W_p (y) \chi_\omega =
\omega^2 \chi_\omega
\label{m7**}
\end{equation}
where $\chi(y)$ must be symmetric across the brane, and
\[
W_p (y) = V(y) + U(y) p^2
\]
(the $\delta $-function term (\ref{Eq/Pg8/1:paper}) is implicit in $V(y)$).
This effective potential has volcano shape, as shown in
figure~\ref{figure3}. At $p=0$, continuum of bulk modes starts at $\omega
= \frac{3}{2} k$, and there is a bound state with $\omega = 0$ whose wave
function is
\begin{equation}
\chi_0 (y) = N\beta^{\frac{3}{2}} (y)
\label{Eq/Pg10/1:version4}
\end{equation}
where $N$ is a normalization factor, ensuring
\[
\int \limits_{0}^{\infty }|\chi _0(y)|^2dy=1
\]
This solution corresponds to $\phi_0 (y) = \mbox{const}$. We assume
that this is the only bound state at $p=0$.

By continuity, at $p \ll k$ the bound state is still present.
Depending on the shapes of $U(y)$ and $V(y)$, there may or may not
exist quasi-localized modes at $p\sim k$. For sufficiently small
width of the ``potential'' $U(y)$ (see (\ref{m6*})),
at $p \gg k$ even
quasi-localized modes cease to exist, while continuum of bulk
modes starts at $\omega^2 = \epsilon^2 p^2 + \frac{9}{4}k^2$.
These properties are illustrated in Appendix.

As an example, the localized solution to equations~(\ref{8*}) and (\ref{8+})
has the form
\[
  \chi_0 = N\mbox{exp}\left[-\left(\frac{3}{2} k - \frac{p^2}{3k}\right)
y \right]
\]
and exists only for
\[
  p < \frac{3}{\sqrt{2}} k
\]
At $p \ll k$ the dispersion relation is $\omega^2 = p^2 [1+ O(\epsilon^2)]$
which explains the coefficient in equation~(\ref{8**}).

\section{Inflation with time-independent parameters}
\label{timedependent}

\subsection{Preliminaries}

At large negative times, the third term in equation~(\ref{7}) dominates,
so all modes are shifted in the bulk towards large $y$, where
$U(y) = \epsilon^2$. For given $p$, all frequencies tend to
$\omega = \epsilon p$ as $\eta \to -\infty$. Thus, at large negative
times the evolution proceeds in the bulk and is adiabatic; the field
$\chi$ has well defined decomposition into positive- and
negative-frequency components,
\begin{equation}
\chi = \int~\frac{d\lambda}{ \sqrt{2\pi\omega_{\lambda,p}}}
\left[ \chi_{\lambda,p} \mbox{exp} \left(-i
\int~\omega_{\lambda,p} d\eta
\right) A^-_{\lambda,p} + \mbox{h.c.} \right]
\label{3.11+}
\end{equation}
where $\omega_{\lambda,p}^2 (\eta)$ and $\chi_{\lambda,p}(y;\eta)$
are eigenvalues and properly normalized eigenstates of the operator
\begin{equation}
  U(y)p^2 + a^2(\eta) [-\partial_y^2 + V(y)]
\label{3.11*}
\end{equation}
and $A_{\lambda,p}^{\pm}$ are creation and annihilation operators. The
asymptotics of $\chi_{\lambda,p}$ at large $y$ is
\[
  \chi_{\lambda,p} \to \mbox{cos} (\lambda y +
\varphi_{\lambda,p})
\]
where the phase $\varphi_{\lambda,p}$ depends on $\eta$. Thus,
\begin{equation}
\omega_{\lambda,p}^2 (\eta) = a^2 (\eta)
\left(\lambda^2 + \frac{9}{4} k^2 \right) + \epsilon^2 p^2
\label{11*}
\end{equation}

As the physical momenta
\[
   P(\eta) = \frac{p}{a(\eta)}
\]
get redshifted, the properties of modes change. At late times
the third term in equation~(\ref{7}) is negligible for superhorizon modes
($P \ll H$), and besides the bulk modes, there is a localized mode
\begin{equation}
\chi_{\mathrm{loc}} (y,\eta) = N\beta^{\frac{3}{2}} (y) \psi_0
(\eta)
\label{3.12+}
\end{equation}
where $\psi_0(\eta)$ obeys the usual equation for superhorizon modes
in four dimensions,
\[
 -\ddot{\psi}_0 + \frac{\ddot{a}}{a} \psi_0 = 0
\]
and has a growing component
\begin{equation}
\psi_0 = \phi _0\cdot a(\eta)
\label{3.12*}
\end{equation}
where $\phi _0$ is independent of time.
This is the standard superhorizon mode. It is localized on the
brane, the wave function in the transverse dimension being given by
equation~(\ref{Eq/Pg10/1:version4}). In terms of the original field $\phi $,
we have for the localized part
\[
\phi _\mathrm{loc}=\phi _0
\]
Our goal is to relate the constant $\phi _0$ in equation~(\ref{3.12*}) to
the creation and annihilation operators entering equation~(\ref{3.11+}).
This analysis is different for $k \gg H$ and $k \lesssim H$.

\subsection{$k \gg H$}
\label{sec:large k}

It is straightforward to see that in the case $k \gg H$, the evolution
of the scalar field is adiabatic when the physical momentum
$P(\eta)$ becomes of order $k$, since at that time the physical
frequencies are high,
\[
\Omega (\eta) \equiv \frac{\omega(\eta)}{a(\eta)} \gg H
\]
and slowly varying,
\[
 \frac{|\dot{\Omega}|}{a\Omega^2} \lesssim \frac{H}{k}
\]
At the time when $P \sim k$, a bound state with $\Omega \sim k \gg H$
emerges in the spectrum of the operator (\ref{3.11*}), and then
its frequency
$\omega (\eta)$ slowly decreases from $\omega \sim ak$ at the time when
$P \sim k$ to
$\omega = p$ at later times when $P \ll k$ but still $P \gg H$.
In this time interval the evolution is still adiabatic.
Since there is a gap of order $k$ between positive- and negative-frequency
parts of the spectrum,  there is no mixing between
the positive- and negative-frequency modes at the time when the bound
state appears, so the bound state is initially in the adiabatic vacuum,
up to corrections suppressed by $H/k$.
Therefore, its further evolution proceeds in the same way as in the
standard four-dimensional scenario; at the horizon crossing, $P \sim H$,
the shape of the bound state is already given by equation~(\ref{3.12+})
modulo $H/k$-corrections. It is important here that by the time of the
horizon crossing, there is no cross talk between the bound state and bulk
modes, since they are also separated by the gap of order $k$ (in terms of
physical frequencies).

We conclude that for $k \gg H$, the spectrum and amplitude of
perturbations coincide with the standard predictions up to
corrections suppressed by $H/k$.

\subsection{$ k \lesssim H$}
\label{sec:low k}

In this case the physical momentum crosses $H$ at the same time as
it crosses $k$,
or earlier. Adiabaticity gets violated at the same time or before
the bound state appears.
This case is hard to treat in detail (see, however,
sections~\ref{sec:timedependent} and~\ref{sec:example}), but for pure
de~Sitter inflation, equation~(\ref{1.1+}), there is a simple scaling
argument showing that the spectrum in the localized mode is flat. Let us
introduce a new time variable
\[
\zeta = p \eta
\]
Equation (\ref{7}) written in terms of $\zeta$ does not contain $p$
explicitly. The initial condition is
\[
\chi \propto \frac{1}{\sqrt{\epsilon p}}
\mbox{e}^{-i \epsilon \zeta} \int~d\lambda
(\chi_\lambda (y,\zeta) A_{\lambda, p}^{-} + \mbox{h.c.})
\]
where the factor $(\epsilon p)^{-1/2}$ comes from
$(\omega_{\lambda,p})^{-1/2}$ in equation~(\ref{3.11+}), and we recalled
that $\omega_{\lambda,p} = \epsilon p$ as $\eta \to -\infty$. Thus, up to
overall factor $p^{-1/2}$, the initial condition does not contain $p$
 explicitly either. The boundary condition (\ref{print7*}) is independent
of $p$. So, the solution to equation~(\ref{7}) has the general form
\[
\chi = \frac{1}{\sqrt{p}} F(\zeta)
\]
where for given $\zeta$, the operator $F$ is a linear combination
of $A_{\lambda,p}^{\pm}$ with $p$-independent coefficients.
Therefore, towards the end of inflation, the localized part of the
solution, for subhorizon three-momenta, behaves as follows (see
equation~(\ref{3.12+})),
\[
\chi _\mathrm{loc}=\psi _0(\eta )\chi _0(y)
\]
where $\chi _0(y)$ is given by equation~(\ref{Eq/Pg10/1:version4}) and
\begin{eqnarray}
\psi _0(\eta ) &=& \frac{1}{\sqrt{p}} \frac{1}{\zeta} \hat{O}_{\bf p}
\nonumber \\
&=& \frac{1}{p^{3/2}} \frac{1}{\eta} \hat{O}_{\bf p}
\nonumber
\end{eqnarray}
where $\hat{O}_{\bf p}$ is a time-independent linear combination of
$A^{\pm}_{\lambda,p}$ with $p$-independent coefficients. The latter
property implies that $\hat{O}_{\bf p}$ describes a Gaussian field obeying
\[
\langle \hat{O}_{\bf p} \hat{O}_{\bf p^{\prime}} \rangle = C \delta ^3({\bf
p} - {\bf p^{\prime}})
\]
where $C$ is independent of $p$. Recalling the relation (\ref{1.2*}), we
obtain that the  localized modes
\[
\phi_0 ({\bf p})\equiv \psi _0(\mathbf{ p})a^{-1}(\eta )
\]
towards the end of inflation make up a Gaussian field with flat
spectrum,
\begin{equation}
\langle \phi_0 ({\bf p}) \phi_0 ({\bf p^{\prime}}) \rangle =
\frac{\mbox{const}}{p^3} \delta^3 ({\bf p} - {\bf p^{\prime}})
\label{15+}
\end{equation}
It is clear that this relation holds only for momenta
which are superhorizon by the end of inflation, higher momenta being
suppressed in the usual way. Therefore, if the overall amplitude of
perturbations is much smaller than 1, their back reaction on the
background geometry is likely to be small. The latter remark applies also
to more interesting cases studied in section~\ref{sec:timedependent}.

It is also clear that for $k \lesssim H$, the constant in equation~(\ref{15+})
depends on $k$ and $H$ in a non-trivial way (see also
sections~\ref{sec:timedependent} and \ref{sec:example}). The latter
observation implies, in particular, that for non-de~Sitter inflation, the
expression for the tilt does not generaly coincide with that in the
standard scenario.

\section{Inflation with slowly varying $k$}
\label{sec:timedependent}

\subsection{Decaying $k$}
\label{decaying k}

\begin{figure}[htb]
\unitlength=1mm
\begin{center}
\begin{picture}(155.00,70.00)(-30,5)
\put(8,57){\mbox{$k$}}
\put(39,8){\mbox{$\eta_\times$}}
\put(85,8){\mbox{$\eta$}}
\put(8,45){\mbox{$H$}}
\includegraphics[height=70mm]{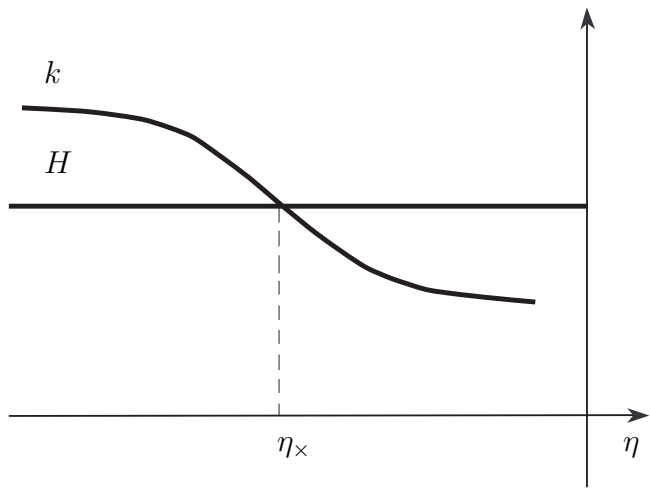}
\end{picture}
\end{center}
\caption{\label{figure4} Decaying $k$.}
\end{figure}

Let us now consider inflation with time-dependent bulk parameters.
As an example, let us assume that $k$ slowly decreases, while, for
definiteness, $H$ stays constant at inflation. Let $k$ cross $H$ at some
$\eta = \eta_{\times}$, as shown in figure~\ref{figure4}. We assume in
what follows that $k$ is not grossly different from $H$ after
$\eta_\times$, and will not count factors of $H/k$. On the other hand, we
consider $\epsilon$ as a small parameter, and will be interested in the
dependence of perturbations on this parameter. The analysis of modes with
low and high three-momenta gives different results.

(i) {\bf Low three-momenta}. These are modes that cross out the de~Sitter
horizon before the time $\eta_\times$, i.e.,
\begin{equation}
   P(\eta_\times) \equiv \frac{p}{a(\eta_\times)} \ll H
\label{4.16*}
\end{equation}
The analysis of these modes goes through as in section~\ref{sec:large k},
the spectrum is flat and the amplitude of perturbations in the
localized mode is given by the standard four-dimensional formula
\begin{equation}
  \phi_0 (p)
 \sim  \frac{H}{p^{3/2}}
\label{standard}
\end{equation}

(ii) {\bf High three-momenta}. These modes obey
\begin{equation}
\epsilon P(\eta_\times) \gg H
\label{4.16**}
\end{equation}
We note in passing that there is an intermediate range of three-momenta,
such that $H \ll P(\eta_\times) \ll H \epsilon^{-1}$. This range is actually
wide, since we assume $\epsilon \ll 1$.

High-momentum modes (\ref{4.16**}) are adiabatic at $\eta = \eta_\times$.
We will see  that the interesting range of transverse momenta
is $\lambda \ll k$. Then the interesting modes
get out of the adiabatic regime at the time $\eta_\epsilon$ when
\begin{equation}
\epsilon P(\eta_\epsilon) \sim H
\label{17*}
\end{equation}
(we call this moment ``$\epsilon$-horizon crossing'';
note that this moment occurs after $k$ crosses $H$,
i.e., $\eta_\epsilon > \eta_\times$). At this
time $P(\eta_\epsilon) \gg k$, so for $\lambda \lesssim k$ the modes are still
far away from the brane, and the frequencies are all of order $\epsilon p$.
Just before the $\epsilon$-horizon crossing, the frequencies (\ref{11*})
are
\begin{equation}
\omega_{\lambda,p} = \epsilon p
\sqrt{1+\frac{\lambda^2 + \frac{9}{4}k^2}{H^2}}
\label{17**}
\end{equation}
where we made use of equation~(\ref{17*}) to write
\[
a(\eta_\epsilon) = \frac{\epsilon p}{H}
\]
In what follows we will often
omit the second factor in equation~(\ref{17**}),
since it is of order~1.

The frequency (\ref{17**}) is of order $a(\eta_\epsilon)H$, which means
that adiabaticity indeed just becomes violated. According to
equation~(\ref{3.11+}), the amplitude of $\chi$ at this moment for all
interesting $\lambda$ is
\begin{eqnarray}
\chi(\eta_\epsilon) &\propto& \frac{1}{\sqrt{\omega}}
\nonumber \\
&\propto& \frac{1}{(\epsilon p)^{1/2}}
\nonumber
\end{eqnarray}
During some time after $\epsilon$-horizon crossing, the modes are still
away from the brane, so equation~(\ref{7}) may be approximated as (modulo
mixing\footnote{This mixing can be shown not to change the estimates that
follow, provided that $k$ is not grossly different from $H$, which we
assume throughout.} between modes with different $\lambda$)
\[
-\ddot{\chi}_\lambda + \frac{2}{\eta^2} \chi_\lambda - \frac{1}{\eta^2}
\frac{\lambda^2 + \frac{9}{4}k^2}{H^2}\chi_\lambda = 0
\]
For $\lambda \ll k$ (and $k<H$, recall that $\eta_\epsilon >
\eta_\times$), this equation has both decaying and growing solutions. The
growing solution, which is of primary interest, behaves as follows,
\begin{equation}
\chi_\lambda \propto \eta^{-\alpha(\lambda) + 1/2}
\label{growth}
\end{equation}
with
\[
\alpha (\lambda) = \sqrt{\frac{9}{4} - \frac{\lambda^2 +
\frac{9}{4}k^2}{H^2}}
\]
This behaviour terminates at the time $\eta_k$ when
\begin{equation}
P(\eta_k) \equiv \frac{p}{a(\eta_k)} \sim k
\label{beta}
\end{equation}
Around this time the wave functions change considerably; in particular,
the localized mode emerges.

The  amplitude of $\chi$ at this moment is
\begin{eqnarray}
\chi_\lambda (\eta_k) &\sim& \chi(\eta_\epsilon)
\left( \frac{\eta_\epsilon}{\eta_k}\right)^{\alpha (\lambda) - 1/2}
\nonumber \\
&\sim& \frac{1}{(\epsilon p)^{1/2}} \left(
\frac{H}{\epsilon k}\right)^{\alpha (\lambda)-1/2}
\nonumber
\end{eqnarray}
With $\epsilon \ll 1$,
the integral over the transverse momenta $\lambda$
is saturated at $\lambda \ll k$, which
justifies the assumption made earlier.
Since all relevant parameters ($P$, $k$ and $H$) are
roughly of the same order of magnitude at this time,
the estimate of the amplitude of the localized mode
at $\eta \sim \eta_k$ is, up to logarithms,
\[
\chi_\mathrm{loc} (\eta_k) \sim \frac{1}{\sqrt{p}}
\frac{1}{\epsilon^{\alpha}} f\left( \frac{H}{k} \right)\cdot \chi _0(y )
\]
where $\chi _0(y)$ is still given by equation~(\ref{Eq/Pg10/1:version4}),
\[
\alpha \equiv \alpha(\lambda = 0) =\frac{3}{2}
\sqrt{1 - \frac{k^2}{H^2}}
\]
and
$f(H/k)$ is roughly of order 1. Let us note that for very small
three-momenta, the value of $k$ entering here is the asymptotic
value at $\eta \gg \eta_\times$.

Later on, at $\eta > \eta_k$, the localized
mode grows as $a(\eta)$ (since $k<H$, this mode is superhorizon
at $\eta_k$ and later). Hence, towards the end of inflation
the amplitude of the localized mode is (we use the notation (\ref{3.12+}))
\[
\psi _0 (\eta) \sim \frac{1}{\sqrt{p}}\frac{1}{\epsilon^{ \alpha}}
\frac{a(\eta)}{a(\eta_k)} f\left( \frac{H}{k} \right)
\]
Recalling equation~(\ref{beta}), we
obtain in terms of the original field $\phi$ that the localized
mode of perturbations stays frozen, and its amplitude is
\begin{equation}
\phi_0 (p) = \frac{H}{p^{3/2}} \frac{1}{\epsilon^{ \alpha}}
\tilde{f}\left( \frac{H}{k} \right)
\label{result}
\end{equation}
where $\tilde{f}$ is again roughly of order 1.
We obtain that at high momenta obeying the relation~(\ref{4.16**}), the
spectrum is again flat, while the amplitude is enhanced by the factor
\begin{equation}
\epsilon^{ -\alpha }
\label{inc}
\end{equation}
as compared to the modes in the low momentum region,
equation~(\ref{standard}). Note that $0<\alpha \leq 3/2$ with $\alpha \to
3/2$ as $k/H\to 0$.

The overall behaviour of perturbations is shown in figure~\ref{figure1}.
Let us make several remarks concerning this result. First, the enhancement
of the amplitude, as given by~(\ref{inc}), is a free parameter,
and can be of several orders of magnitude. The reason for this enhancement
is clear: modes with low three-momentum become ``tachyonic'' when they are
still living in the bulk. For $\epsilon =0$ and constant $k<H$,  all modes
are tachyonic ``from the beginning'', the situation reminiscent of that in
four-dimensional theory with the Corley--Jacobson dispersion
relation~\cite{d1,d2}. With our setup, we effectively cut off the
divergence associated with the tachyonic instability. Second, the
beginning of the transition region $p_\times $ is determined by the time
$\eta_\times$ when $k$ crosses $H$. Clearly, the physically interesting
situation occurs when this time is not very distant from the time
inflation ends. Third, the momenta in the end and in the beginning of the
transition region differ at least by a factor $\epsilon^{-1}$ (this is the
case shown in figure~\ref{figure1}), but their ratio may be much larger,
since it is not necessarily determined by the parameter $\epsilon$ only,
as one might guess by comparing the relations~(\ref{4.16*}) and
(\ref{4.16**}). The value of the parameter $k$ entering the exponent
$\alpha$ is to be evaluated somewhere in the interval $(\eta_\epsilon (p),
\eta_k(p))$, so for very slowly varying $k(\eta)$ this value in fact
depends on three-momentum $p$ and settles down only as $k(\eta_k(p))$
reaches its asymptotic value. Fourth, the parameter $\epsilon$ may itself
depend on time, which would lead to even more complicated spectrum of
perturbations. Finally, although we have considered an example with $k$
crossing $H$, an interesting spectrum is obtained also if $k<H$ during
entire inflationary epoch, provided that $k$ and/or $\epsilon $ depend on
time. The latter two points are clear from equation~(\ref{result}), which
is valid for $k<H$, and the above observation that $\epsilon $ and $k$
entering equation~(\ref{result}) depend on three momenta, as they are to
be evaluated at some $\eta $ belonging to the interval $(\eta _\epsilon
(p),\eta _k(p))$.

\begin{figure}[t]
\unitlength=1mm
\begin{center}
\begin{picture}(155.00,70.00)(-30,5)
\put(8,30){\mbox{$k$}}
\put(39,8){\mbox{$\eta_\times$}}
\put(85,8){\mbox{$\eta$}}
\put(8,45){\mbox{$H$}}
\includegraphics[height=70mm]{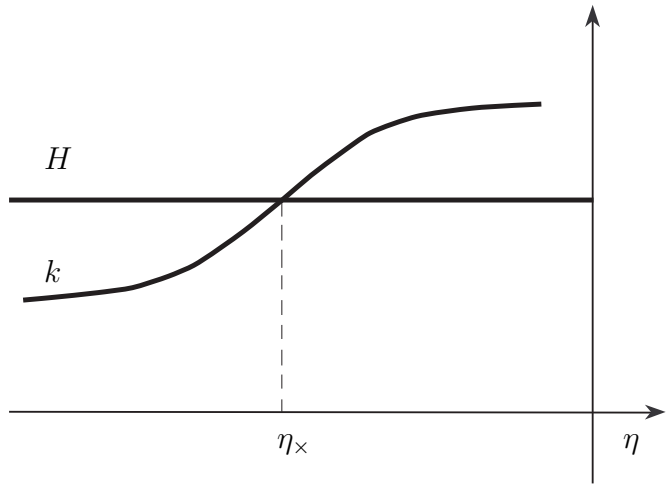}
\end{picture}
\end{center}
\caption{\label{figure5} Increasing $k$.}
\end{figure}

\subsection{Increasing k}

If $k$ increases in time as shown in figure~\ref{figure5},
the properties of
soft and hard modes (in three-dimensional sense)  are reversed.
{\bf Soft modes} in this case
are those for which the bound state gets formed before
$k$ crosses $H$, i.e.
\[
P(\eta_\times) \equiv \frac{p}{a(\eta_\times)} \ll k
\]
It is these modes that become tachyonic when still living
in the bulk.
They increase as given by equation~(\ref{growth}) from the time
$\eta_\epsilon$ to $\eta_k$; these times are still
determined
by equations~(\ref{17*}) and (\ref{beta}), with $k$ now being
the value of $k(\eta)$ in the past, where $k<H$.
At the time
$\eta_k$ the bound state emerges, and then the amplitude of
 the bound state mode increases as $a(\eta)^{-1}$. Thus,
equation~(\ref{result}) applies now to the soft modes.

{\bf Hard modes} are those which never become tachyonic
when staying in the bulk, which is the case for
\[
 \epsilon P(\eta_\times) \gg H
\]
They start growing after forming the bound state, so
their amplitudes are given by the four-dimensional expression
(\ref{standard}). The overall spectrum is thus the reverse
of figure~\ref{figure1}; it is shown in figure~\ref{figure2}.
The remarks made in the end of section~\ref{decaying k}
apply to the case of increasing $k$ as well.

\section{Explicit example}
\label{sec:example}

In this section we consider an explicit setup outlined in the end of
section~\ref{sec:setup}. The field $\chi $ obeys equation~(\ref{8*}) with
the boundary condition (\ref{8+}). We will be interested in the case
$k\sim H$ and, as in section~\ref{sec:timedependent}, study the dependence
of the amplitudes on $\epsilon $, assuming $\epsilon \ll1$. As  mentioned
in section~\ref{sec:setup}, in spite of the fact that variables separate
in equation~(\ref{8*}), the dynamics is non-trivial due to the
time-dependence of the boundary condition (\ref{8+}). We simplify the
analysis further by making use of the instantaneous approximation to the
boundary condition (\ref{8+}). Namely, at $\eta <\eta _k$, where $\eta _k$
is given by equation~(\ref{beta}), i.e.
\begin{equation}
\eta _k=-\frac{k}{Hp}\sim -\frac{1}{p}
\label{Eq/Pg20/1A:paper}
\end{equation}
the third term in equation~(\ref{8+}) dominates and we omit the first two
terms. So, at these times the boundary condition reads
\begin{equation}
\chi |_{y=0}=0 \ \ \mathrm{ at} \ \ \eta <\eta _k
\label{Eq/Pg19/1:paper}
\end{equation}
On the other hand, at $\eta >\eta _k$ we neglect the third term in
equation~(\ref{8+}) and the boundary condition becomes
\begin{equation}
\left( \chi '+\frac{3}{2}k\chi \right)_{y=0}=0 \ \ \mathrm{ at}\ \ \eta
>\eta _k
\label{Eq/Pg20/1:paper}
\end{equation}
In this approximation, the problem reduces to that of solving  the field
equation (\ref{8*}) with the boundary conditions
(\ref{Eq/Pg19/1:paper}), (\ref{Eq/Pg20/1:paper}) and matching the
solutions at $\eta =\eta _k$.

At $\eta <\eta _k$ the solution is (we slightly change the notation as
compared to equation~(\ref{3.11+}))
\[
\chi =\int \limits_{0}^{\infty }\frac{d\lambda }{\sqrt{2\pi }}
\left[
\chi
_{\lambda,p} (y,\eta )A_{\lambda ,p}^-+\mathrm{ h.c.}
  \right]
\]
where $\chi _{\lambda ,p}$ are
\begin{equation}
\chi _{\lambda ,p}(y,\chi )=-i\sqrt{\pi }\,\mathrm{ e}^{\frac{i\pi
\alpha }{2}}\sqrt{-\eta }H^{(1)}_\alpha (-\eta \epsilon p)
\sin(\lambda y)
\label{Eq/Pg20/3:paper}
\end{equation}
and
\begin{equation}
\alpha =\sqrt{\frac{9}{4}\left(1-\frac{k^2}{H^2}\right)-\frac{\lambda
^2}{H^2}}
\label{Eq/Pg20/4:paper}
\end{equation}
As $\eta \to -\infty $, these modes approach properly normalized
negative-frequency plane waves
\[
\chi _{\lambda ,p}(y,\eta )=\sqrt{\frac{2}{\omega }}\mathrm{ e}^{-i\omega
\eta +i\varphi }\sin(\lambda y)
\]
where $\omega =\epsilon p$ and $\varphi $ is a constant phase. For $k<H$
and small enough $\lambda $ the parameter $\alpha $ is real. So, these
modes indeed grow like $|\eta|^{1/2-\alpha }$ in the interval $\eta
_\epsilon \lesssim \eta \lesssim \eta _k$, where $\eta _\epsilon $ is
determined by equation~(\ref{17*}). For $k>H$, the parameter $\alpha $ is
imaginary, so the modes (\ref{Eq/Pg20/3:paper}) oscillate in $\eta $
for all $\eta <\eta _k$.

At late times, $\eta >\eta _k$, besides the bulk modes, there are two
localized modes obeying equations~(\ref{8*}) and
(\ref{Eq/Pg20/1:paper}),
\[
\chi _\mathrm{loc}^{\pm}(y,\eta )=\sqrt{3k}\cdot \sqrt{-\eta }(-\eta
)^{\pm\frac{3}{2}}\mathrm{ e}^{-\frac{3}{2}ky}
\]
These are the standard four-dimensional modes; recall that $k\sim H$, so
at $\eta >\eta _k$ these modes are superhorizon. At $\eta >\eta _k$ the
field $\chi $ is a superposition
\begin{equation}
\chi (y,\eta )=C^{(-)}\chi _\mathrm{loc}^-(y,\eta )+C^{(+)}\chi
^{+}_\mathrm{loc}(y,\eta )+ \mbox{bulk modes}
\label{Eq/Pg21/1A:paper}
\end{equation}
The coefficients here are obtained by matching this solution to the
solution (\ref{Eq/Pg20/3:paper}) at $\eta =\eta _k$. We are interested
in the coefficient $C^{(-)}$, which is the amplitude of the growing
localized mode. To this end we notice that the bulk modes entering
equation~(\ref{Eq/Pg21/1A:paper}) are orthogonal to $\chi _\mathrm{
loc}^{\pm}$ with the standard scalar product
\[
(\psi ,\chi )=\int \limits_{0}^{\infty }dy(\dot{\chi }\psi ^*-\dot{\psi
}^*\chi )
\]
On the other hand,
\[
(\chi _\mathrm{loc}^+,\chi _\mathrm{loc}^-)=3
\]
Thus
\[
C^{(-)}=\frac{1}{3}(\chi _\mathrm{loc}^+,\chi )_{\eta =\eta _k}
\]
where $\chi $ is the solution (\ref{Eq/Pg20/3:paper}), and the scalar
product is to be evaluated at $\eta =\eta _k$. One obtains for arbitrary
$\eta $
\[
(\chi _\mathrm{loc}^+,\chi )=\sqrt{3k\pi } \frac{(-\eta
p\epsilon )^{\frac{3}{2}}\mathrm{
e}^{\frac{i\pi \alpha }{2}}}{(p\epsilon)^{\frac{3}{2}
}}\!\!\left(\frac{3}{2}H^{(1)}_\alpha (-\eta \epsilon p)-\eta
\dot{H}^{(1)}_\alpha (-\eta \epsilon p) \right)\!\!\!\frac{i\lambda
}{\lambda ^2+\frac{9}{4}k^2}
\]
Thus
\begin{equation}
C^{(-)}=\frac{1}{\sqrt{3k}p^\frac{3}{2}}\int \limits_{0}^{\infty
}\frac{d\lambda }{\sqrt{2}}\mathrm{ e}^{\frac{i\pi \alpha
}{2}}\!\!\left(\frac{3}{2}H^{(1)}_\alpha (\epsilon )-\epsilon \partial
_\epsilon H^{(1)}_\alpha(\epsilon ) \right)\frac{ik\lambda }{\lambda
^2+\frac{9}{4}k^2}A_{\lambda ,p}^-+\mathrm{ h.c.}
\label{Eq/Pg21/4:paper}
\end{equation}
where we took into account equation (\ref{Eq/Pg20/1A:paper}).

Now we are ready to calculate the spectrum of perturbations in the
localized mode, defined as
\[
\langle \phi _0(\mathbf{ p})\phi _0(\mathbf{ p}')\rangle \equiv
\frac{2\pi ^2}{p^3}\mathcal{ P}(p)\delta ^3(\mathbf{ p}-\mathbf{ p}')
\]
We make use of the relation (\ref{1.2*}) of the original field $\phi $ to
the field $\chi $ and the standard commutation relations
\[
[A_{\lambda ,p}^-,A_{\lambda' ,p'}^+]=\delta ^3(\mathbf{ p}-\mathbf{
p}')\delta (\lambda -\lambda ')
\]
With the expression (\ref{Eq/Pg21/4:paper}) for the coefficient
$C^{(-)}$ we find
\begin{equation}
\mathcal{ P}(p)=\frac{H^2}{4\pi ^2}\int \limits_{0}^{\infty }d\lambda
\mathrm{ e}^{\frac{i\pi (\alpha -\alpha ^*)}{2}}\frac{\lambda
^2k}{3\left(\lambda
^2+\frac{9}{4}k^2\right)^2}\left|\frac{3}{2}H^{(1)}_\alpha (\epsilon
)-\epsilon \partial _\epsilon H_\alpha ^{(1)}(\epsilon )\right|^2
\label{Eq/Pg21/7:paper}
\end{equation}
For time-independent $k$, the right hand side of
equation~(\ref{Eq/Pg21/7:paper}) is independent of $p$, the spectrum is
flat, while the amplitude depends on $k$, $H$ and $\epsilon $ in a
non-trivial way. This is in accord with the general analysis given in
section~\ref{timedependent}.

On the other hand, if $k$ slowly varies in time, the spectrum is no longer
flat. The important time interval is $\eta _\epsilon <\eta <\eta _k$, in
which bulk modes grow provided that $k<H$, and do not grow for $k>H$.

Let us consider the case of decreasing $k$, figure~\ref{figure4}.
For low three-momenta one has (see equation~(\ref{4.16*})) $\eta _\times\gg
\eta _k$ and  $\alpha (k)$ defined in
(\ref{Eq/Pg20/4:paper}) is pure imaginary.
 So, the integrand in equation~(\ref{Eq/Pg21/7:paper}) is independent
of $\epsilon $ at small $\epsilon $. The spectrum is flat, and the power is
\[
\mathcal{ P}(p)=\frac{H^2}{4\pi^2 }\cdot\mathrm{ const},
\]
where the constant is of order 1.
In the opposite case of high three-momenta, the times are ordered as
$\eta _\times<\eta _\epsilon <\eta _k$. So, $k(\eta _k)<H$ and $\alpha
(k,\lambda )$ is real at small enough $\lambda $. In this case
\[
\left|\frac{3}{2}H^{(1)}_\alpha (\epsilon
)-\epsilon \partial _\epsilon H_\alpha ^{(1)}(\epsilon )\right|^2
=\frac{1}{4\pi }\left(\frac{2}{\epsilon }   \right)^{2\alpha }\left(\alpha
+\frac{3}{2} \right)^2\frac{\Gamma (\alpha )^2}{\Gamma (3/2)^2}
\]
Substituting this expression into equation~(\ref{Eq/Pg21/7:paper}) one
finds that the integral is saturated at small transverse momenta, $\lambda
^2\sim H^2/\ln \epsilon $. Therefore, we again obtain the flat spectrum
but with the amplitude enhanced by the factor (\ref{inc}) (up to
logarithm), in agreement with the discussion in section~\ref{decaying k}.

If $k$ increases in time as shown in figure~\ref{figure5}, $\alpha (k)$
is imaginary for high momenta ($\eta _\times>\eta _k$) and real for low
momenta ($\eta _\times<\eta _\epsilon $). The perturbations get enhanced
by the factor $\epsilon ^{-\alpha }$ at low momenta, and do not get
enhanced at high momenta, so we obtain the spectrum shown in
figure~\ref{figure2}.

\ack
The authors are indebted to S.~Dubovsky, M.~Sazhin,
M.~Shaposhnikov, S.~Sibiryakov and W.~Unruh
for helpful discussions and comments. This work is
supported in part by RFBR grant 05-02-17363-a, by the Grants of the
President of Russian Federation NS-2184.2003.2, MK-3507.2004.2, by INTAS
grant YSF 04-83-3015, and by the grant of the Russian Science Support
Foundation.

\section*{Appendix}
In this Appendix we illustrate the properties of eigenmodes
of equation (\ref{m7**}) by considering a step-function ``potential''
\[
U(y) = \alpha^2 \theta (y_0-|y|) + \epsilon^2 \theta (|y| -y_0)
\]
where $\alpha \sim 1$ and $\epsilon \ll 1$. We take the warp factor
$\beta (y) $ in the form (\ref{m7+}), so that another ``potential''
is constant,
\[
  V(y) = \frac{9}{4} k^2
\]
We consider equation (\ref{m7**}) on half-line $y>0$ and impose the
boundary condition (\ref{print7*}), i.e.,
\begin{equation}
\left( \chi^\prime + \frac{3}{2} k
 \chi\right)_{y=0} = 0
\label{mbc}
\end{equation}
We are interested in localized or quasilocalized modes whose
general expression at $y<y_0$ is
\begin{equation}
\chi_\omega = A \mbox{e}^{-\nu y} + B \mbox{e}^{\nu y}
\; , \;\;\;\; y <y_0
\label{ma*}
\end{equation}
Then
\begin{equation}
\omega^2 = -\nu^2 + \frac{9}{4}k^2 + \alpha^2 p^2
\label{mfreq}
\end{equation}
while the boundary condition (\ref{mbc}) reads
\begin{equation}
\left( \frac{3}{2}k - \nu\right) A
+
\left( \frac{3}{2}k + \nu\right) B =0
\label{ma+}
\end{equation}

Let us begin with localized modes. They decay at $y>y_0$,
\[
\chi_\omega = \mbox{e}^{-\lambda (y -y_0)} \; , \;\;\;\;
y>y_0
\]
where we set the overall factor equal to one.  From
equations~(\ref{m7**}) and (\ref{mfreq}) we obtain
\begin{equation}
   \nu^2 = \lambda^2 + \hat{\alpha}^2 p^2
\label{ma++}
\end{equation}
where $\hat{\alpha}^2 = \alpha^2 - \epsilon^2$.
By matching $\chi_\omega$ and $\chi_\omega^{\prime}$
at $y=y_0$ we find
\begin{eqnarray}
A &=& \frac{\nu + \lambda}{2\nu} \mbox{e}^{\nu y_0}
\nonumber \\
B &=& \frac{\nu - \lambda}{2\nu} \mbox{e}^{-\nu y_0}
\nonumber
\end{eqnarray}
and the relation (\ref{ma+}) gives
\begin{equation}
\mbox{e}^{-2\nu y_0} =
\frac{\nu + \lambda}{\nu - \lambda} \cdot
\frac{\nu - \frac{3}{2}k}{\nu + \frac{3}{2}k}
\label{ma+++}
\end{equation}
This equation, supplemented with equation (\ref{ma++}),
determines the eigenvalue $\nu$ for given three-momentum $p$.

At $p^2=0$ equation (\ref{m7**}) has a zero mode
$\chi_0 = \mbox{exp} (-\frac{3}{2}ky)$ with $\lambda = \nu
= \frac{3}{2} k$ and $\omega =0$. For $p^2 \ll k^2$ there exists
a solution to equations (\ref{ma++}) and (\ref{ma+++}) describing a
localized state
close to the zero mode,
\begin{eqnarray}
\nu &=& \frac{3}{2} k + \frac{\hat{\alpha}^2 }{3k}
\mbox{e}^{-3ky_0} \cdot p^2 + O(p^4)
\nonumber \\
\omega^2 &=& \left(\alpha^2 - \hat{\alpha}^2 \mbox{e}^{-3ky_0}
\right)\cdot p^2 +O(p^4)
\nonumber
\end{eqnarray}
Up to rescaling of coordinates on the brane, this mode has relativistic
dispersion relation.

At $p^2 \gg k^2$ no real solutions to equations (\ref{ma++}),
(\ref{ma+++}) exist: equation (\ref{ma++}) gives
$\nu \gtrsim p$, so the right hand side
of equation (\ref{ma+++}) is greater than 1 in that case. Thus,
there are no localized modes at high three-momenta $p$.

Let us now turn to quasilocalized modes. They still have the form
(\ref{ma*}) at $y<y_0$, while at $y>0$ they oscillate
\[
\chi_\omega = b\mbox{e}^{iq(y-y_0)} + b^* \mbox{e}^{-iq(y-y_0)}
\]
so that
\[
   q^2 = \hat{\alpha}^2 p^2 - \nu^2
\]
The normalization condition appropriate for the modes in continuum
is
\begin{equation}
 |b|^2 = \frac{1}{2}
\label{ma***}
\end{equation}
By matching $\chi_\omega$ and $\chi_\omega^{\prime}$
at $y=y_0$ we obtain
\[
b =\frac{1}{2}\left(A\mbox{e}^{-\nu y_0}
+B \mbox{e}^{\nu y_0} \right) + \frac{i\nu}{2q}
\left(A\mbox{e}^{-\nu y_0}
-B \mbox{e}^{\nu y_0} \right)
\]
so the normalization condition (\ref{ma***}) and equation
(\ref{ma+}) determine the coefficient $A$,
\begin{equation}
|A|^2 = 2\left[ \left(\mbox{e}^{-\nu y_0}
+ \frac{\nu - \frac{3}{2}k}{\nu + \frac{3}{2}k}
\mbox{e}^{\nu y_0} \right)^2
+
\frac{\nu^2}{q^2}
\left(\mbox{e}^{-\nu y_0}
- \frac{\nu - \frac{3}{2}k}{\nu + \frac{3}{2}k}
\mbox{e}^{\nu y_0} \right)^2 \right]^{-1}
\label{mastar}
\end{equation}
The defining property of a quasilocalized mode is that it is
large at the brane position (when normalized according to
(\ref{ma***})),
\begin{equation}
|A| \gg 1
\label{ma****}
\end{equation}
At $ky_0 \gg 1$ this relation indeed holds for
\[
\nu = \frac{3}{2} k + O(\mbox{e}^{-3ky_0})
\]
In that case the quasilocalized mode exists for arbitrarily
high three-momenta $p$; furthermore, at $p \gg k$ and
\[
\nu = \frac{3}{2} k - 3k \mbox{e}^{-3ky_0}
+ O\left(\frac{k^2}{p} \mbox{e}^{-3ky_0}\right)
\]
one has
\[
|A| \sim \frac{p}{k} \mbox{e}^{3ky_0/2}
\]
so the localization gets stronger at high momenta $p$.

At $ky_0 \lesssim 1$ the only way the relation (\ref{ma****})
may be satisfied is that $p$ is large, and the first term in
the square bracket in (\ref{mastar}) vanishes, which gives
\[
\mbox{e}^{-2\nu y_0} =
\frac{\frac{3}{2}k - \nu}{\frac{3}{2}k + \nu}
\]
The latter equation, however, has positive solutions only for
\[
y_0 > \frac{2}{3k}
\]
Thus, a quasilocalized mode exists at high three-momenta $p$
provided that $ky_0$ is large enough, otherwise there are no
no quasilocalized modes  at $p \gg k$. It is the latter case
that is considered in the main text.

\end{document}